\documentclass[%
reprint,
superscriptaddress,
nofootinbib,
amsmath,amssymb,
aps,
pra,
showkeys,
]{revtex4-2}

\usepackage[normalem]{ulem} 
\usepackage{url}
\usepackage{xcolor}
\usepackage{hyperref} 
\usepackage[nameinlink,capitalize]{cleveref}
\hypersetup{
	colorlinks = true,
	linkbordercolor = {white},
	linkcolor={purple},
	citecolor={purple},
	urlcolor={blue}
}
\usepackage{comment}
\usepackage{cleveref}

 \usepackage{textcomp}

\usepackage[utf8]{inputenc}
\usepackage{mathtools}
\usepackage{algorithm2e}
\usepackage{physics}
\usepackage{graphicx}
\usepackage{dcolumn}
\usepackage{bm}
\usepackage{verbatim}
\usepackage{xcolor}
\usepackage{subcaption}
\usepackage{multirow}

\usepackage{ mathrsfs }

\usepackage{amsmath}

\definecolor{forestgreen}{rgb}{0.13, 0.55, 0.13}

\usepackage[labelfont=bf,
   justification=Justified,
   format=plain]{caption}

\begin{document}

\title{Detecting Markovianity of Quantum Processes via Recurrent Neural Networks}

\author{Angela Rosy Morgillo{\href{https://orcid.org/0009-0006-6142-0692}{\includegraphics[scale=0.004]{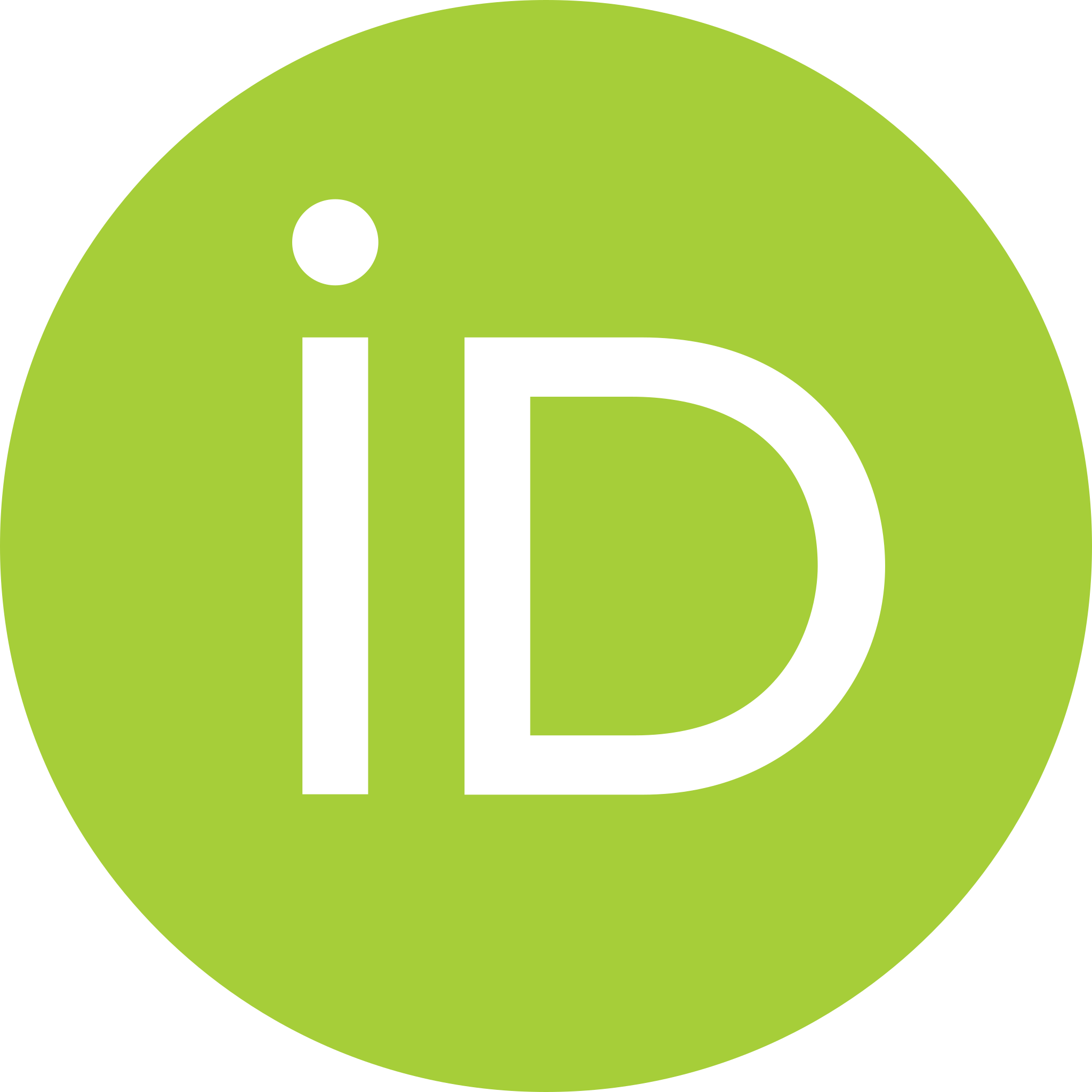}}}}
\email{angelarosy.morgillo01@universitadipavia.it}
\affiliation{Dipartimento di Fisica, Università di Pavia, Via Bassi 6, I-27100, Pavia, Italy}
\affiliation{INFN Sezione di Pavia, Via Bassi 6, I-27100, Pavia, Italy}

\author{Massimiliano F. Sacchi{\href{https://orcid.org/0000-0002-8909-2196}{\includegraphics[scale=0.004]{images/orcid.png}}}}
\affiliation{CNR-Istituto di Fotonica e Nanotecnologie, Piazza Leonardo da Vinci 32, I-20133, Milano, Italy}
\affiliation{Dipartimento di Fisica, Università di Pavia, Via Bassi 6, I-27100, Pavia, Italy}

\author{Chiara Macchiavello{\href{https://orcid.org/0000-0002-2955-8759}{\includegraphics[scale=0.004]{images/orcid.png}}}}
\affiliation{Dipartimento di Fisica, Università di Pavia, Via Bassi 6, I-27100, Pavia, Italy}
\affiliation{INFN Sezione di Pavia, Via Bassi 6, I-27100, Pavia, Italy}

\date{\today}

\begin{abstract}
We present a novel methodology utilizing Recurrent Neural Networks (RNNs) to classify Markovian and non-Markovian quantum processes, leveraging time series data derived from Choi states. The model exhibits exceptional accuracy, surpassing 95\%, across diverse scenarios, including dephasing and Pauli channels in an arbitrary basis, generalized amplitude damping dynamics, and even in the presence of noise. Additionally, the developed model shows efficient forecasting capabilities for the analyzed time series data. These results suggest the potential of RNNs in discerning and predicting the Markovian nature of quantum processes.
\end{abstract}

\maketitle

\section{\label{sec:level1}INTRODUCTION}
The study of open quantum dynamics plays a crucial role in understanding the behaviour of quantum systems interacting with their surrounding environment~\cite{rivas2012open, banerjee2018open}. Unlike closed quantum systems, which are isolated from external influences, open quantum systems experience continuous exchanges of energy, information, and particles with their surroundings. Such interactions lead to the phenomenon of quantum decoherence, where the system loses its coherence and exhibits classical-like behaviour~\cite{schlosshauer2019quantum}.

One key aspect in characterizing the dynamics of open quantum systems is the concept of Markovianity. Markovian processes are stochastic processes where the future state of the system depends only on its present state and not on the sequence of events that preceded it~\cite{rivas2014quantum}. In the theory of open quantum systems, Markovian dynamics provides a powerful framework for describing the time evolution of reduced open system states. This evolution is condensed in a quantum master equation, typically expressed in the Lindblad form, in which a weak system-environment coupling is assumed~\cite{lindblad1976generators, MERKLI2020167996, Merkli2022dynamicsofopen}. 

Nonetheless, in nature, the Markovian assumption often breaks down, especially in scenarios characterized by strong system-environment coupling and low temperatures, leading to the emergence of non-Markovian dynamics and information backflow~\cite{hall2014canonical, shrikant2023quantum, utagi2020temporal}.  

\begin{figure*}[t]
    \centering
    \includegraphics[trim={0.5cm 14.8cm 1.02cm 6.7cm},clip,width=1\textwidth]{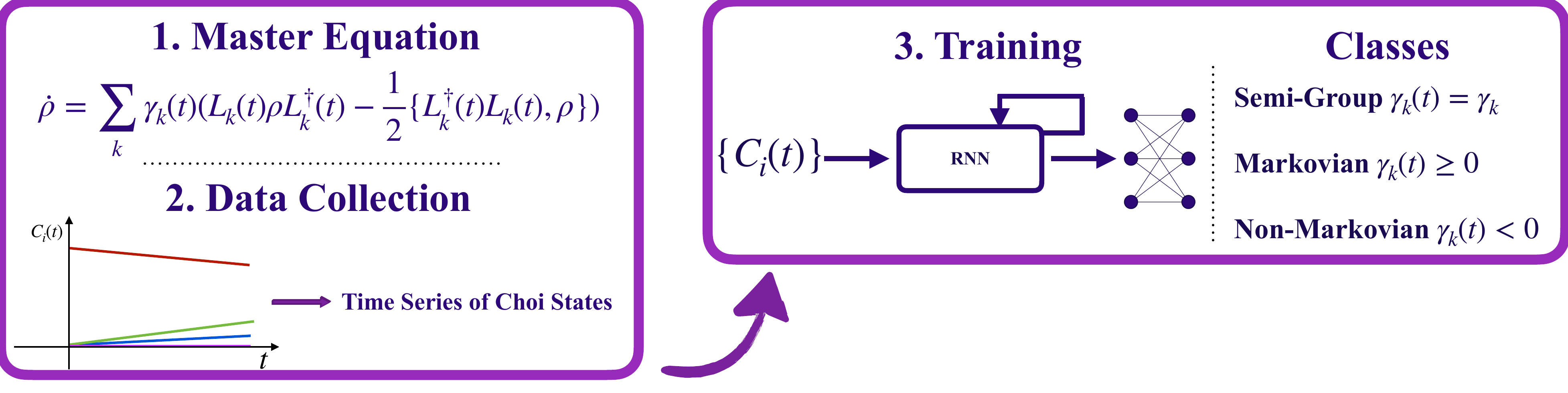}
    \caption{Our model scheme. We start with the formulation of a master equation to depict the system evolution. We then derive the diagonal components $C_i(t)$ of the Choi states in the Bell basis at different time steps, elucidating the time evolution process outlined in the master equation. The resulting dataset, including time series inputs and class labels (semi-group, Markovian and non-Markovian), is then fed into a recurrent neural network, that outputs the probability distribution over the three classes.}
    \label{fig:Scheme}
\end{figure*}

In recent years, distinguishing Markovian and non-Markovian dynamics in open quantum systems has gained substantial interest in various domains such as quantum information theory~\cite{bylicka2014non, cheong2022communication, liu2013nonunital}, quantum control~\cite{reich2015exploiting}, system-environment coupling modeling~\cite{li2018concepts}, quantum error mitigation~\cite{takagi2022fundamental, hakoshima2021relationship}, and quantum thermodynamics~\cite{pmlr-v162-mutti22a, choquehuanca2023non}, in which non-Markovian dynamics has been shown to facilitate faster energy conversion of heat and work~\cite{huang2022strong}.\\
At the same time, machine and deep learning have recently achieved notable results, particularly in computer vision~\cite{chai2021deep} and natural language processing~\cite{khurana2023natural, min2023recent}, with transformer models revolutionizing language understanding~\cite{deng2022benefits}. \\
Indeed, machine learning has shown great potential also in quantum error mitigation~\cite{kim2020quantum}, entanglement quantification~\cite{gray2018machine, koutny2023deep} and classification~\cite{harney2021mixed}, and in quantum thermodynamics for optimal control of finite time processes in quantum thermal machines~\cite{khait2022optimal}, for predicting open quantum dynamics described by time-local generators~\cite{mazza2021machine}, for reducing entropy production in closed quantum systems out of equilibrium~\cite{sgroi2021reinforcement}, for classifying the environmental parameters of single-qubit dephasing channels using a time series approach~\cite{palmieri2021multiclass}, for classifying the spectral density characterizing the dynamics of a system~\cite{barr2023spectral}, for modeling non-Markovian effects in several regimes by using Recurrent Neural Networks~\cite{banchi2018modelling} and for identifying Pareto-optimal cycles~\cite{erdman2023pareto}.\\
In this study, we present a model based on Recurrent Neural Networks (RNNs)~\cite{salehinejad2017recent}, tailored for analyzing time series data representing Choi states of quantum channels~\cite{choi1975completely, jamiolkowski1972linear}. A significant characteristic of our methodology is the utilization of only the diagonal components within the Bell basis of the Choi state, resulting in results comparable to those achieved when considering all components (see \Cref{fig:acc} for more details). Our model is also effective in classifying quantum processes in the presence of noise, maintaining high accuracy even under noisy conditions. The novelty of our approach lies in using machine learning for the classification task, as, to the best of our knowledge, machine learning has only been applied to assess the degree of non-Markovianity~\cite{fanchini2021estimating}. Unlike current state-of-the-art methods, which rely on analytical solutions and are therefore limited to specific classes of dynamical maps with explicitly solvable master equations, our machine learning-based approach provides greater flexibility by handling datasets for processes with dynamics that are analytically intractable or computationally unresolvable. The schematic representation of our approach is illustrated in \Cref{fig:Scheme}.

Indeed, prior studies have demonstrated the effectiveness of RNNs, along with other machine learning techniques such as artificial neural networks and support vector machines, in discerning various external noise sources, as illustrated in~\cite{martina2023machine}. Furthermore, the Choi state has emerged as a valuable resource for assessing non-Markovianity in quantum processes, as exemplified in~\cite{chruscinski2017detecting, mallick2023assessing}.

Our primary aim is to tackle the classification problem of discerning between semi-group, Markovian, and non-Markovian dynamics. Here, a semi-group denotes a specific type of Markovian dynamics characterized by constant Lindblad operators and decay rates in the canonical master equation. On the other hand, (non-)Markovianity refers to the case of (non-)positive time-dependent decay rates. We apply this approach to various single-qubit channels, including dephasing, Pauli and generalized amplitude damping. \\
Our model achieves accuracies exceeding 95\% and demonstrates strong forecasting capabilities, providing robust and reliable predictions.
\\
The manuscript is organized as follows. In \Cref{sec:theory} we introduce the dynamics of open quantum systems, defining the concept of (non-)Markovianity. \Cref{Sec:models} introduces the quantum models employed in constructing the temporal sequences essential to our study. Moving to \Cref{sec:methods}, we show the methodologies applied, specifically introducing RNNs and time series and describing the construction of the datasets. \Cref{sec:results} presents the results achieved from process classification and forecasting, showing the utilized network architecture. Finally, in \Cref{sec:conclusion} we derive our conclusions and outline promising paths for future research.

\section{Theory}
\label{sec:theory}
Exploring the dynamics of open quantum systems is a key focus in modern quantum physics, holding implications for quantum information processing and technology. In this section we investigate the fundamental aspects of open quantum dynamics, starting from the definition of the master equation, followed by the characterization of dynamical maps and their alternative representation using Choi states. Subsequently, we define the concept of quantum (non-)Markovianity.

\subsection{Dynamics of Open Quantum Systems}

\paragraph*{The Master Equation.---}
Quantum systems are inherently susceptible to decoherence and dissipation when interacting with their surroundings. We use master equations to understand and model these effects systematically. 
The dynamics of a system coupled to an environment is described by the time non-local Nakajima-Zwanzig master equation~\cite{rivas2012open}
\begin{equation}
    \dot{\rho}(t) = -\frac{i}{\hbar} [H_S, \rho(t)] + \int_{t_0}^t \mathcal{K}_{t,s}[\rho(s)] ds,
    \label{eq:master_equation_kernel}
\end{equation}
where $H_S$ is the system Hamiltonian and the memory kernel $\mathcal{K}_{t,s}$ is a linear map representing the effects of the environment on the system. Utilizing the Born-Markov and secular approximations allows us to approximate the memory kernel, eliminating rapidly fluctuating terms and yielding a memoryless time-local Lindblad master equation:
\begin{equation}
    \dot{\rho} =
-\frac{i}{\hbar} [H_S, \rho] + \sum_k \gamma_k \Big(A_k\rho A^{\dagger}_k - \frac{1}{2}\{A_k^{\dagger}  A_k, \rho\} \Big),
\label{eq:master_equation_not}
\end{equation}
where $\{A_k\}$ represent the Lindblad operators, and the decay rates $\gamma_k$ are positive coefficients~\cite{hall2014canonical}.

We can generalize the master equation in the canonical form with time-dependent decay rates, where the rotating approximation holds while the Born-Markov approximation does not~\cite{shrikant2023quantum}:
\begin{align}
    \dot{\rho} &= \mathscr{L}(t)[\rho] = -\frac{i}{\hbar} [H_S(t), \rho]  \notag\\
    &+\sum_k \gamma_k(t) \Big(A_k(t)\rho A^{\dagger}_k (t) - \frac{1}{2}\{A_k^{\dagger} (t) A_k(t), \rho\} \Big).
    \label{eq:master_equation}
\end{align}
The canonical form of the master equation requires that the operators $A_k(t)$ constitute a set of orthonormal traceless operators~\cite{hall2014canonical}.
This form proves particularly valuable for characterizing non-Markovianity, as discussed in~\Cref{sec:MNM}, defining a \textit{unique} set of decoherence rates.

\paragraph*{Dynamical Maps and Choi States.---}
The solution of \Cref{eq:master_equation} provides a set of dynamical maps $\Lambda(t,t_0)$, transforming density operators $\rho$ into density operators $\Lambda(t, t_0)[\rho]$. In order to be physically admissible, the maps $\Lambda(t, t_0)$ have to be completely positive (CP) for all times $t$~\cite{petrucbreue}. These maps can be represented as
\begin{equation}
    \Lambda(t,t_0)[\rho] = \sum_k E_k(t) \rho E_k^\dagger(t),
\end{equation}
where $\{E_k(t)\}$ denotes the set of Kraus operators, satisfying $\sum_k E_k^\dagger (t) E_k (t) =I$~\cite{nielsen2010quantum}. From linearity, $\Lambda(t,t_0)$ itself satisfies $\dot{\Lambda}(t, t_0)=\mathscr{L}(t)[\Lambda(t, t_0)]$, and its solution takes the form:
\begin{equation}
    \Lambda(t,t_0) = \mathcal{T}e^{\int_{t_0}^t\mathscr{L}(s) ds},
\end{equation}
with $\mathcal{T}$ representing the time-ordering operator.

For each dynamical map, a one-to-one correspondence can be established with a bipartite state through the Choi-Jamiołkowski isomorphism~\cite{adlam2020operational}. Given a completely positive dynamical operator $\Lambda(t,t_0)$, the associated Choi-Jamiołkowski state $\mathcal{C}_{\Lambda}(t, t_0)$ is derived by applying $\Lambda(t,t_0)$ to a party of a maximally entangled bipartite state $\dyad{\phi^+}$~\cite{mallick2023assessing}:
\begin{equation}
    \mathcal{C}_{\Lambda} (t, t_0) = ( \Lambda(t,t_0) \otimes I) \dyad{\phi^+}.
    \label{eq:choiformula}
\end{equation}
For single-qubit states, we can select the Bell state $|\phi^+\rangle = \frac{1}{\sqrt{2}}(|00\rangle + |11\rangle)$, and represent the Choi state on the Bell basis $\{|\phi^{\pm}\rangle = \frac{1}{\sqrt{2}}(|00\rangle \pm |11\rangle), |\psi^{\pm}\rangle = \frac{1}{\sqrt{2}}(|01\rangle \pm |10\rangle)\}$.

\subsection{Markovian vs Non-Markovian}
\label{sec:MNM}
A time-local master equation describes a Markovian process if and only if the canonical decay rates reported in~\Cref{eq:master_equation} are positive at all times $t$. Conversely, if any of the coefficients are negative, we refer to the dynamics as non-Markovian.
In our study, we examine three types of processes, namely \textit{semi-group}, \textit{Markovian} and \textit{non-Markovian}, distinguished by the behaviour of the decay rates $\gamma_k(t)$ in \Cref{eq:master_equation}. 

Markovianity is established when $\gamma_k(t) \geq 0 \, \forall \,k$ for all times $t$, suggesting a memoryless evolution. Here, Markovianity is equivalent to the property of CP-divisibility of the dynamical map~\cite{rivas2014quantum, hall2014canonical}. However, the resulting set of dynamical maps $\Lambda(t)$ generally does not constitute a semi-group i.e. they do not satisfy the semi-group property $\Lambda(t+s) = \Lambda(t) \circ \Lambda(s)$ for all $t,s \geq 0$. Specifically, we refer to a \textit{semi-group} type for dynamical maps when all coefficients $\gamma_k(t)$ are constant and positive, and the operators $H_S(t)$ and $A_k(t)$ are time-independent, representing a particular instance of Markovianity~\cite{lidar2001completely}. Finally, if one of the coefficients $\gamma_k(t)$ in \Cref{eq:master_equation} is negative for certain times $t$, non-Markovianity arises.

\section{\label{Sec:models} Quantum Processes}
In this section, we provide an extensive overview of the various quantum processes under investigation. These encompass the dephasing and the Pauli channels in arbitrary basis, and the generalized amplitude damping channel. Our exploration includes a detailed examination of each channel, showing in particular the relation between the decay rates of the canonical master equation and the corresponding parameters characterizing the quantum channels.

First, we exclusively consider pure dissipative evolutions, thus disregarding the Hamiltonian $H_S(t)$ in \Cref{eq:master_equation}. While this is not strictly necessary for the methodology, focusing on pure dissipative processes simplifies the implementation and is particularly relevant to our context. Including the coherent part of the process (i.e., the Hamiltonian) would still allow the method to remain valid. Second, we restrict our analysis to channels where all Lindblad operators $A_k(t)$ are time-independent, facilitating the analytical derivation of the Choi states by simplifying the resolution of the master equations.  If the Lindblad operators were time-dependent, the semigroup class would no longer be applicable, as it requires constant decoherence coefficients and time-independent Lindblad operators. However, the methodology could be extended to consider time-dependent operators by focusing on the classification of Markovian versus non-Markovian processes.

In the upcoming paragraphs, we will utilize the notation $\mathcal{L}_A[\cdot] \coloneq A\cdot A^\dagger - \frac{1}{2}\{A^\dagger A, \cdot\}$. 

\subsection{Dephasing Channel in a Rotated Basis}
We begin by formulating the master equation that governs the dynamics of a dephasing channel:
\begin{equation}
   \dot{\rho}(t) = \frac{\gamma (t)}{2} \mathcal{L}_{\sigma_z} [\rho(t)],
   \label{eq:deph}
\end{equation}
where $\gamma(t)$ denotes the decay rate and $\sigma_z = |0\rangle\langle 0| - |1\rangle \langle1|$ is the Pauli matrix associated with the $z-$axis.

The solution to this differential equation is  
\begin{equation}
\rho(t)= \frac{1+e^{-\Gamma(t)}}{2}\rho(0) + \frac{1-e^{-\Gamma(t)}}{2} \sigma_z\rho(0)\sigma_z 
\label{eq:soldeph}
\end{equation}
with $\Gamma(t) = \int_0^t\gamma(s) ds$.  

The Kraus operators characterizing the dephasing channel are then given by:
\begin{equation}
    K_1(t) = \sqrt{\frac{1+e^{-\Gamma(t)}}{2}} I, \quad  K_2(t) = \sqrt{\frac{1-e^{-\Gamma(t)}}{2}}\sigma_z,
    \label{eq:krausdeph}
\end{equation}
where $I = |0\rangle \langle 0| + |1\rangle \langle 1|$.
The Choi state of the dephasing (D) channel is obtained by the evolution of the maximally entangled state $|\phi^+\rangle$, and can be represented in terms of the Bell basis as follows
\begin{equation}
\mathcal{C}_{\text{D}} (t) =\hspace{-0.007em} \frac{1+e^{-\Gamma(t)}}{2} \hspace{-0.06em} \dyad{\phi^+}\hspace{-0.04em} +\hspace{-0.016em} \frac{1-e^{-\Gamma(t)}}{2} \hspace{-0.06em} \dyad{\phi^-}\hspace{-0.2em}.
\label{eq:choid}
\end{equation}
We can extend this analysis to a dephasing channel with respect to an arbitrary basis, by considering a master equation of the form:
\begin{equation}
    \dot{\rho}(t) = \frac{\gamma (t)}{2}\mathcal{L}_{\bm{\sigma} \cdot \bm{n}} [\rho(t)],
  \label{eq:dephgen}
\end{equation}
with $\bm{\sigma} = (\sigma_x, \sigma_y, \sigma_z)$ being the vector of the three Pauli matrices and $\bm{n} = (\sin\theta \cos\phi, \sin\theta \sin\varphi, \cos\theta)$ being a vector on the unit sphere, with $\theta \in [0, \pi]$ and $\varphi \in [0, 2\pi)$.\\
This scenario is more intricate compared to the standard dephasing channel. While the dynamics of the dephasing process can be effectively analysed and classified using a feedforward neural network, addressing dephasing in a rotated basis calls for a more advanced architecture, such as a recurrent neural network.\\
The Choi state for the dephasing channel in a rotated basis (D$_{\text{RB}}$) is given by:
\begin{widetext}
\begin{align}
  \mathcal{C}_{\text{D}_{\text{RB}}}(t) &=\frac{1 + e^{-\Gamma(t)}}{2}  \dyad{\phi^+} + \frac{1 - e^{-\Gamma(t)}}{2} \{ \cos^{2}{\theta}\dyad{\phi^-}+ \sin^2{\theta}\cos^2{\varphi} \dyad{\psi^+} + \sin^2{\theta} \sin^{2}{\varphi} \dyad{\psi^-} \nonumber \\
  &\quad+ \sin{\theta}\cos{\theta} \cos{\varphi} \big( \dyad{\phi^-}{\psi^+} + \dyad{\psi^+}{\phi^-} \big) + i\sin{\theta}\cos{\theta} \sin{\varphi} \big( \dyad{\phi^-}{\psi^-} - \dyad{\psi^-}{\phi^-} \big)  \nonumber \\
  &\quad+ i\sin^2{\theta}\sin{\varphi}\cos{\varphi} \big( \dyad{\psi^+}{\psi^-} -\dyad{\psi^-}{\psi^+} \big) \}.
  \label{eq:gRB}
\end{align}
\end{widetext}

\subsection{Pauli Channel}

The dynamics of a Pauli channel is described by the master equation:
\begin{equation}
    \dot{\rho}(t) = \frac{1}{2}\sum_{k=1}^3\gamma_k (t)\mathcal{L}_{\sigma_k} [\rho(t)],
\label{eq:genpa}
\end{equation}
where $\sigma_k \in \{\sigma_1, \sigma_2, \sigma_3\}$ represent the Pauli matrices, and $\gamma_k(t)$ are time-dependent coefficients. The generators $\mathcal{L}_{\sigma_k} (\cdot)= \sigma_k \cdot \sigma_k - \cdot$ satisfy the commutation relation $[\mathcal{L}_{\sigma_i}, \mathcal{L}_{\sigma_j}] = 0 \,\, \forall i,j=1,2,3$~\cite{chruscinski2016generalized}. This property allows for the solution to be derived by factorizing the solutions for the dephasing channel~\eqref{eq:soldeph} for all Pauli matrices.

The Choi state for the Pauli channel (P) is given by:
\begin{widetext}
\begin{equation}
\begin{split}
  \mathcal{C}_{\text{P}}(t) = &\frac{1}{4} \{ (1 + e^{-\Gamma_1(t)-\Gamma_2(t)} + e^{-\Gamma_1(t)-\Gamma_3(t)} + e^{-\Gamma_2(t)-\Gamma_3(t)} ) \dyad{\phi^+}  \\
  &+ (1 + e^{-\Gamma_1(t)-\Gamma_2(t)} - e^{-\Gamma_1(t)-\Gamma_3(t)} - e^{-\Gamma_2(t)-\Gamma_3(t)} ) \dyad{\phi^-} \\
  &+ (1 - e^{-\Gamma_1(t)-\Gamma_2(t)} - e^{-\Gamma_1(t)-\Gamma_3(t)} + e^{-\Gamma_2(t)-\Gamma_3(t)} ) \dyad{\psi^+}\\
  &+ (1 - e^{-\Gamma_1(t)-\Gamma_2(t)} + e^{-\Gamma_1(t)-\Gamma_3(t)} - e^{-\Gamma_2(t)-\Gamma_3(t)} ) \dyad{\psi^-} \},
\end{split}
\label{eq:choigp}
\end{equation}
\end{widetext}
where $\Gamma_k(t) = \int_0^t \gamma_k(s) ds$.

\paragraph*{Rotated Basis Transformation.---}The Pauli channel can be generalized by considering a rotated basis, akin to the treatment applied to the dephasing channel. In this scenario, the modification of the Choi state presented in \Cref{eq:choigp} is achieved by accounting for the rotation of the reference system. This is done by considering the total rotation $R(\alpha, \beta, \gamma)=R_z(\gamma)R_y(\beta)R_z(\alpha)$, where $\alpha$, $\beta$ and $\gamma$ are the Euler angles, and $R_i (\psi)$ represents a rotation around the $i$-axis by an angle $\psi$, with $i \in \{x, y ,z\}$.
\subsection{Generalized Amplitude Damping Channel}
The Generalized Amplitude Damping evolution is characterized by the following master equation
\begin{equation}
    \dot{\rho}(t) = \big(\gamma_1(t) \mathcal{L}_{\sigma_-} + \gamma_2(t) \mathcal{L}_{\sigma_+}\big)[\rho(t)].
\end{equation}
Here, $\sigma_+ = |1 \rangle \langle 0|$ and $\sigma_- = |0 \rangle \langle 1|$ denote the raising and lowering operators.

The corresponding channel can be written as $\rho(t) = \sum_{i=0}^3 E_i \rho(0) E_i^\dagger$, where 
\begin{align}
    E_0 &= \sqrt{p}\begin{pmatrix}
        1 & 0 \\
        0 & \sqrt{1-\lambda}
    \end{pmatrix} \quad E_2= \sqrt{1-p}\begin{pmatrix}
        \sqrt{1-\lambda} & 0 \\
        0 & 1
    \end{pmatrix} \nonumber \\ E_1 &= \sqrt{p}\begin{pmatrix}
        0 & \sqrt{\lambda} \\
        0 & 0
    \end{pmatrix} \quad \quad \,\,\,\, E_3= \sqrt{1-p}\begin{pmatrix}
        0 & 0 \\
        \sqrt{\lambda} & 0
    \end{pmatrix},
\end{align}
with $p, \lambda \in [0,1]$. Here, $p+q=1$, and $p$ and $\lambda$ are time-dependent functions which are related to the canonical decay rates as follows~\cite{shrikant2022eternal}:
\begin{equation}
    \gamma_1(t) = \lambda \dot{p} + \frac{\dot{\lambda}p}{1-\lambda} \,\,\,\,\,\,\, \gamma_2(t) = \lambda\dot{q} + \frac{\dot{\lambda}q}{1-\lambda}.
    \label{eq:diffgad}
\end{equation}

The Choi state characterizing the Generalized Amplitude Damping (GAD) channel takes the form:

\begin{widetext}
\begin{equation}
\begin{split}
  \mathcal{C}_{\text{GAD}}(t) = &\frac{(1 + \sqrt{1 - \lambda})^2}{4} \dyad{\phi^+} + \frac{(1 -\sqrt{1 - \lambda})^2}{4} \dyad{\phi^-} + \frac{\lambda}{4} (\dyad{\psi^+} + \dyad{\psi^-} + \dyad{\phi^+}{\phi^-}  \\
    &+ \dyad{\phi^-}{\phi^+} + i(1-2p) \dyad{\psi^+}{\psi^-}- i(1-2p) \dyad{\psi^-}{\psi^+}).
\end{split}
\label{eq:gad}
\end{equation}
\end{widetext}
We observe that the diagonal components of the Choi state are unaffected by $p$.  Hence, our classification between Markovian and non-Markovian behaviour relies solely on the damping parameter $\lambda$.

\section{Methods}
\label{sec:methods}
In this section, we rigorously define the classification problem under consideration – the differentiation between semi-group, Markovian and non-Markovian dynamics. Our approach involves the analysis of the evolution of the Choi states across various single qubit channels.
 
We first start by introducing key concepts related to time series data and recurrent neural networks (RNNs). Then, we discuss the neural network methodology applied in this work, including details about the optimization procedure for our classification problems and the construction of the dataset.

\subsection{Time Series and Recurrent Neural Networks}

Time series data play a fundamental role in various disciplines, allowing to understand and predict dynamic processes in science, economics, and various other fields. A time series is a collection of observations or measurements ordered chronologically, often at equally spaced intervals. This temporal structure enables the analysis of patterns, trends, and fluctuations over time, providing valuable insights into the dynamics of diverse phenomena~\cite{shumway2000time}.

A time series is defined by a sequence of observations $x_t$, each corresponding to a specific time point $t$. In the context of discrete time, which is the focus of our investigation, a discrete-time time series is characterized by observations made at distinct, discrete instances. Formally, this involves a set $\mathscr{T}$ representing the discrete times at which observations are recorded, typically occurring at fixed time intervals~\cite{brockwell2002introduction}.

Sequential data, like time series, can be effectively processed using recurrent neural networks (RNNs), a class of bidirectional artificial neural networks with recurrent connections. 
The recurrent connections create a form of memory, allowing to consider information from past time steps when processing the current input.

Two specific variants of RNNs that tackle challenges associated with long-term dependencies and vanishing gradients are Long Short-Term Memory (LSTM) and Gated Recurrent Unit (GRU). Further details about these architectures can be found in \Cref{app:rnns}.

\subsection{RNNs for Quantum Processes Classification}
In this section, we clarify the application of RNNs in tackling the classification problem of discerning the Markovian nature of quantum processes. 
The analysed dataset comprises time series consisting of selected components of Choi states of the quantum process recorded at different time steps. Each entry is associated with a corresponding label indicating whether the process exhibits semi-group, Markovian or non-Markovian behaviour.

The dataset is structured as follows:
\begin{equation}
\mathcal{D} = {(\bm{\mathcal{X}}^{(i)}, y^{(i)})}_{i=1}^{N},
\end{equation}
where $\bm{\mathcal{X}}^{(i)} = [{\bm{\mathcal{C}}^{(i)}(t_0), \ldots, \bm{\mathcal{C}}^{(i)}(T)}]$ represents the i-th sample, corresponding to a time series of components of the Choi state with time steps $t \in [t_0, T]$. Additionally, $y^{(i)} \in \{\text{0, 1, 2}\}$ denotes the correct class label, where `0' indicates semi-group samples,  `1' signifies Markovian samples, and  `2' represents non-Markovian samples. \\
In our simulations, the time series comprise 7 equally-spaced time steps ranging from 0 to $T=3$ s. The dataset comprises 7200 training samples, 900 validation samples, and 900 test samples.

Through the utilization of a series of recurrent and feed-forward layers applied to the time series samples, the performance of the network is assessed using accuracy as the figure of merit. Accuracy is defined as the ratio of correct predictions to the total number of predictions. The categorical cross-entropy, used as the loss function, is formulated as follows:
\begin{equation}
\label{eq:cat}
\text{CCE} \coloneqq -\sum_{i=1}^{N} \sum_{j=1}^{3} y_{ij} \log(p_{ij}).
\end{equation}
Here, \(N\) denotes the number of samples to be classified, while $3$ represents the number of distinct classes. The true class label $y^{(i)}$ of the $i$-th sample is one-hot encoded in a probability distribution \(\bm{y}_{i}
\in \{0,1\}^3\), where one-hot encoding is a method of representing categorical variables as binary vectors. In this encoding, the true class of the sample is represented by a vector with a 1 in the position of the correct class and 0s in all other positions.

The terms \(y_{ij}\) and \(p_{ij}\), both constrained within the range [0, 1], respectively stand for the actual and model-predicted probabilities for the \(i\)-th sample to belong to the \(j\)-th class. The negative logarithm of the predicted probability is summed over all samples and classes to compute the categorical cross-entropy loss. 

In the next section, we analyse the network efficiency in distinguishing among the classes.

\section{Results}
\label{sec:results}
We now present the results obtained by using RNNs for time series classification and forecasting. We start discussing the performance of our model in classifying whether a process is a semi-group, Markovian or non-Markovian. This evaluation is done for dephasing and Pauli channels in arbitrary bases, and generalized amplitude damping channels. Then, we show also the network capability in predicting the evolution of several components of the Choi state under both dephasing and Pauli dynamics. Our results demonstrate an optimal accuracy in classification and efficient forecasting capabilities for the analysed time series data.

Additional information regarding the data generation process is available in \Cref{app:datageneration}.

All simulations are run with TensorFlow~\cite{tensorflow2015-whitepaper}, with the Adam optimizer~\cite{kingma2014adam} employed. 

\subsection{Channel Classification}
In this section, we present our findings on classifying channels based on the Markovianity of quantum processes. Our model architectures have demonstrated remarkable performance in achieving robust classification accuracy for quantum processes. The neural network architecture employed consists of a GRU with 32 units, augmented by a feedforward neural network comprising 16 neurons, with Rectified Linear Unit (ReLU) as the activation function\footnote{In this specific setup, Feed Forward Neural Networks (FFNNs) achieve comparable accuracies to RNNs. However, RNNs could provide advantages in capturing sequential dependencies and are expected to outperform FFNNs in tasks involving more complex temporal structures, longer time series, or noisy data.}. The architecture culminates with a softmax layer comprising three neurons.

For these simulations, the learning rate is set to 0.001, the number of epochs to 800, and the batch size to 720. The learning processes exhibit noise; however, stability can be attained by enlarging the dataset (approximately 24000 samples may suffice) and reducing the learning rate. Importantly, our model demonstrates effectiveness even with a limited number of samples. 

Notably, both GRU and LSTM architectures are found to be interchangeable. All the tests are conducted exclusively using the diagonal components within the Bell basis of the Choi states as input.

\begin{figure*}
    \centering
    \subfloat[\label{subfig:a1}]{{\includegraphics[trim={0.6cm 0.6cm 0.2cm 0.8cm},clip, width=0.45 \textwidth]{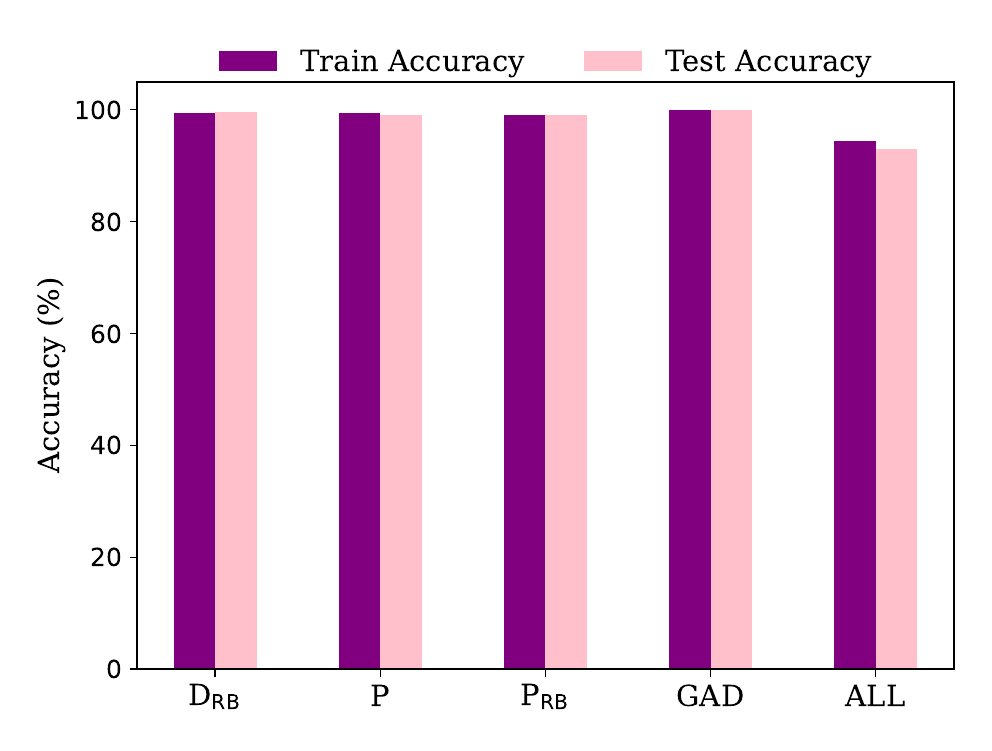} }}%
    \subfloat[\label{subfig:a2}]{{\includegraphics[trim={0.6cm 0.6cm 0.2cm 0.8cm},clip, width=0.45 \textwidth]{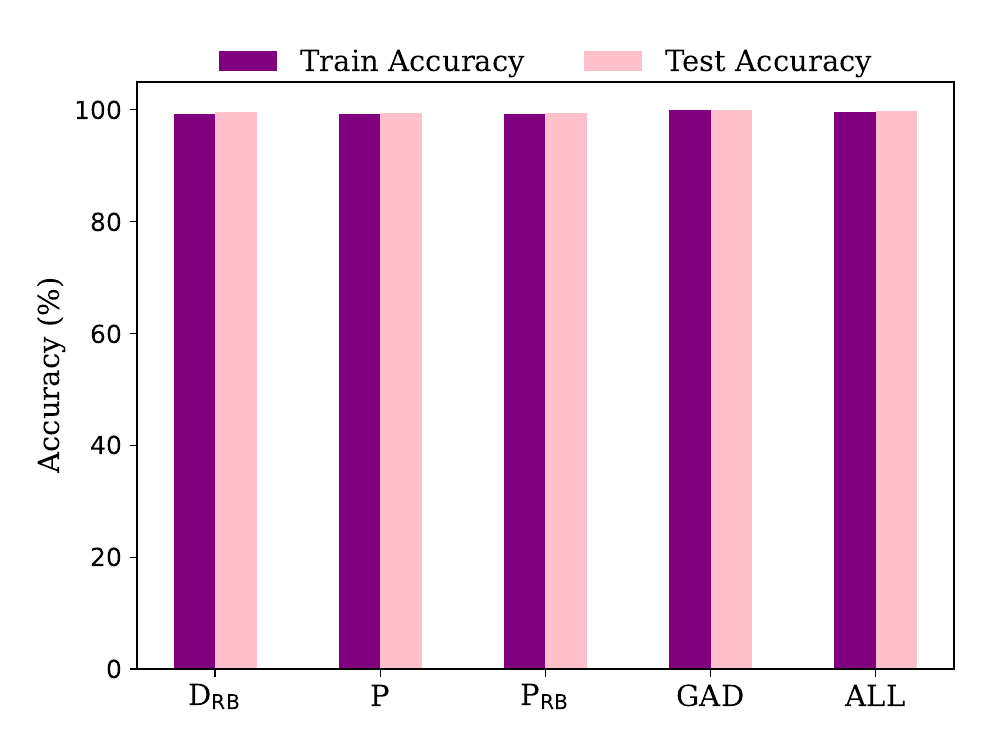} }}\\
    \subfloat[\label{subfig:a3}]{{\includegraphics[trim={0.6cm 0.6cm 0.2cm 0.8cm},clip, width=0.45 \textwidth]{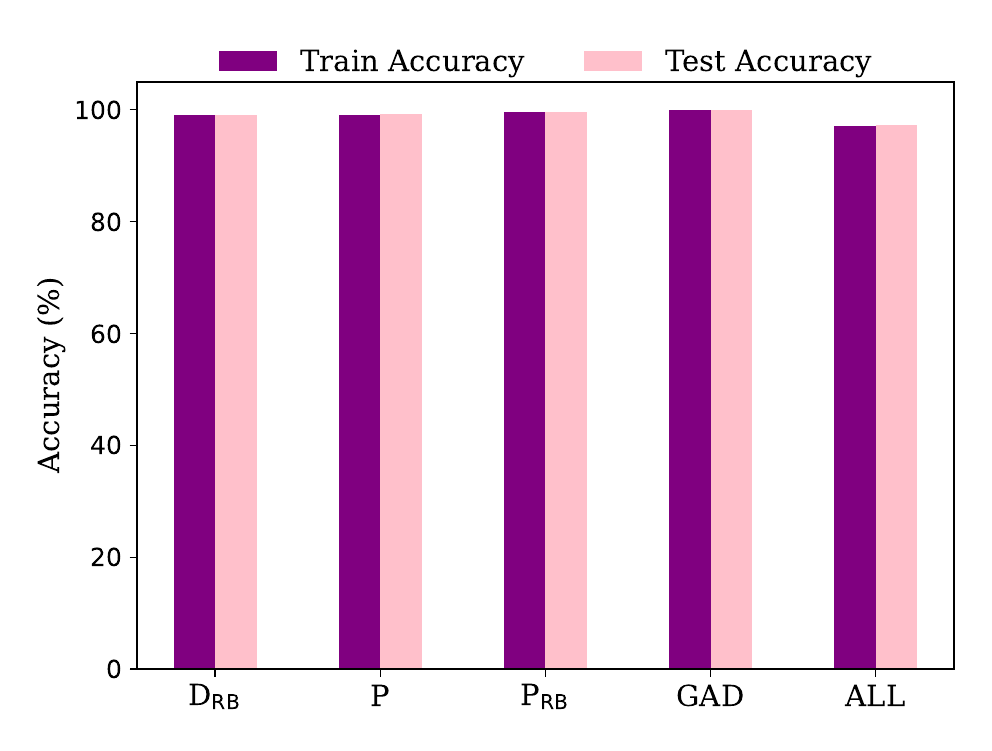} }}%
    \subfloat[\label{subfig:a4}]{{\includegraphics[trim={0.6cm 0.6cm 0.2cm 0.8cm},clip, width=0.45 \textwidth]{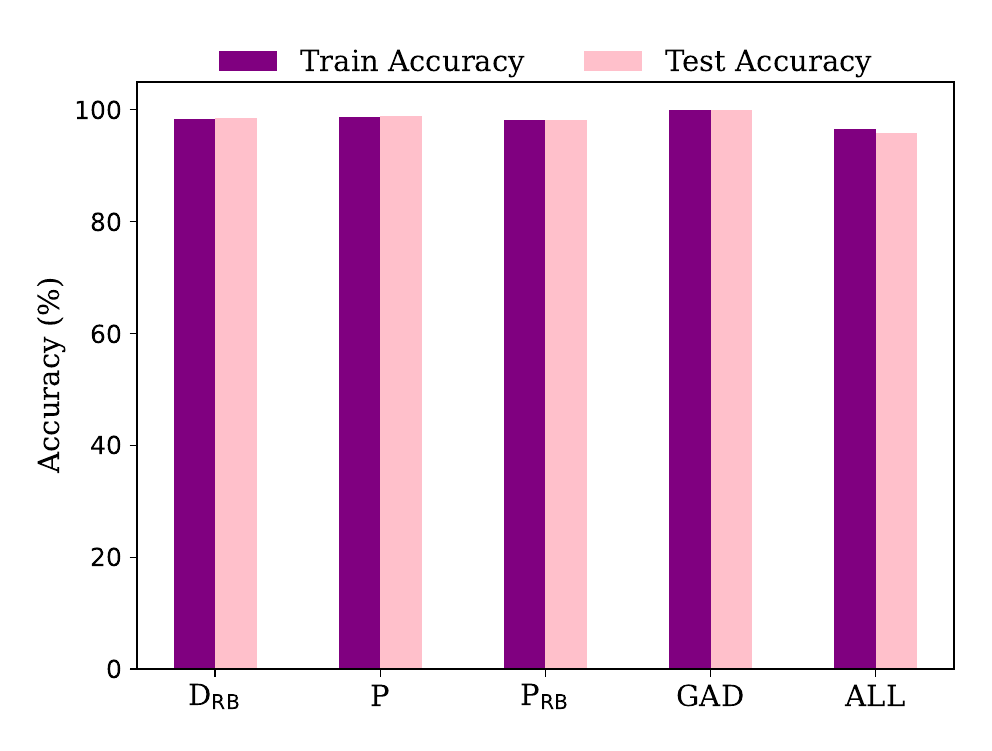} }}%
    \caption{Train and test accuracies are depicted for various processes in a classification problem: (a) when only the diagonal components of the Choi matrix are used, (b) when all the non-null and unique components of the Choi matrix are used, (c) when the Choi states are prepared with a fixed fidelity $F=0.95$ across all samples, and (d) when the Choi states are prepared with fixed fidelity $F=0.5$. These include dephasing in a rotated basis (D$_{\text{RB}}$), Pauli (P), Pauli in a rotated basis (P$_{\text{RB}}$), generalized amplitude damping (GAD) channels, and all the aforementioned channels collectively labeled as ALL.}
    \label{fig:acc}
\end{figure*}

In \Cref{fig:acc}, the resulting training and test accuracies of classification tasks involving the analysed models in \Cref{Sec:models} for single-qubit channels—specifically, the dephasing channel in a rotated basis, the Pauli channel, the Pauli channel in a rotated basis and the generalized amplitude damping channel— are illustrated. We also examine the scenario where all channels are included in the training set. \\
The results presented correspond to a single representative run for each case. The findings have been verified to be robust across multiple independent runs with varying random initializations of the RNN parameters. In particular, both the accuracy and the convergence behavior of the loss function remain consistent, highlighting the reliability of the approach.
We also demonstrate the model's performance when utilizing all components of the Choi matrix, achieving comparable results (see \Cref{subfig:a2}). Additionally, we analyze two distinct noise scenarios: the first involves Choi states prepared with a fixed fidelity of 0.95 across all samples, while the second presents a more challenging scenario where the fidelity of the Choi states is set to 0.5 for each sample. These scenarios are illustrated in \Cref{subfig:a3} and \Cref{subfig:a4}, respectively. See \Cref{app:noise} for more details about the noise employed. In particular, when \( F = 0.25 \), the network predictions reduce to random guessing. This is expected, as \( F = 0.25 \) corresponds to a maximally mixed state, meaning that all information about the channel is lost.\\
Notably, all models examined achieve a test accuracy of at least 95\%, indicating no overfitting, i.e. the overspecialization of the model on the training samples.  
    
\begin{figure}[h!]
    \centering
    \includegraphics[trim={0.35cm 0.05cm 0.4cm 0.0cm},clip,width=0.48\textwidth]{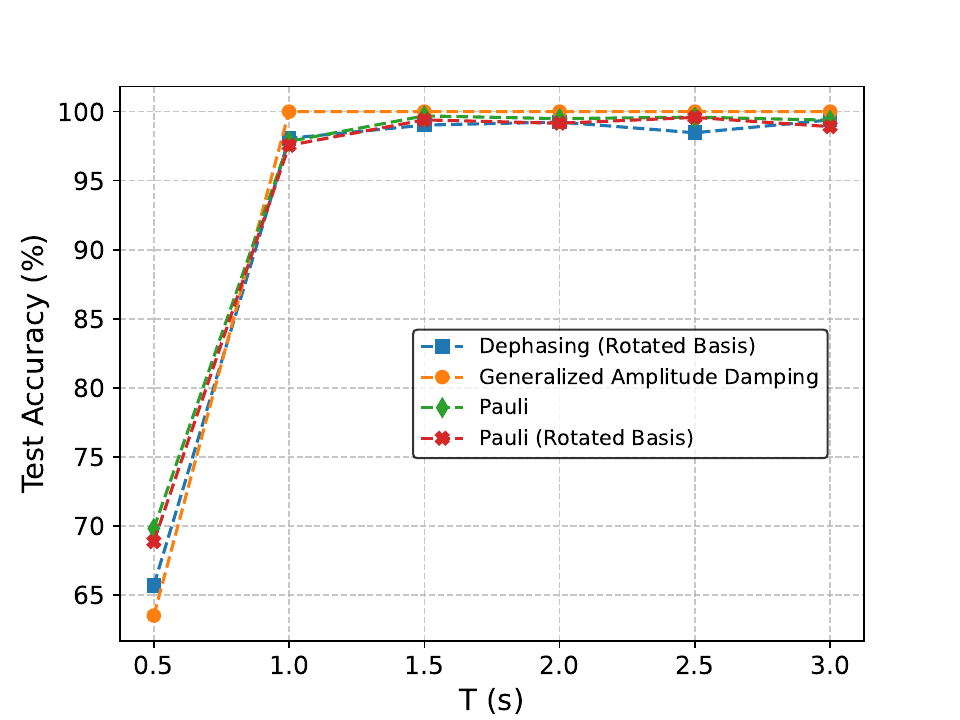}
    \caption{Correlation between classification accuracy and the length of the time series for dephasing, generalized amplitude damping, Pauli and Pauli in a rotated basis channels, with a time step $\Delta t = 0.5$. The $x$-axis represents the final time-step $T$ of the time series, corresponding to different time series lengths. Each reported result is an average across five distinct runs.}
    \label{fig:timeseries}
\end{figure}

\paragraph*{Time Series Length.---}We examine how varying time series lengths affect the classification accuracy of our model across all single-qubit channels previously analysed. The aim is to explore the model sensitivity to the temporal dimension of input data. Remarkably, our findings reveal that, even with a brief time series consisting of only three time steps, our model achieves accuracies exceeding 90\%, as illustrated in \Cref{fig:timeseries}. Each reported result represents an average across five distinct runs. Observed fluctuations may stem from variations in the training process. The consistency of the hyperparameters of the network (epochs $= 800$, batch size $= 600$, learning rate $= 0.001$), along with a fixed training cardinality of 6000 samples across all simulations, suggests that a specific combination of these hyperparameters may be more suitable for a particular time series length. These hyperparameters, however, are specific to this simulation and tailored for this particular task.

\begin{figure*}
    \centering
    \subfloat[\label{subfig:c1}]{{\includegraphics[width=4.3cm]{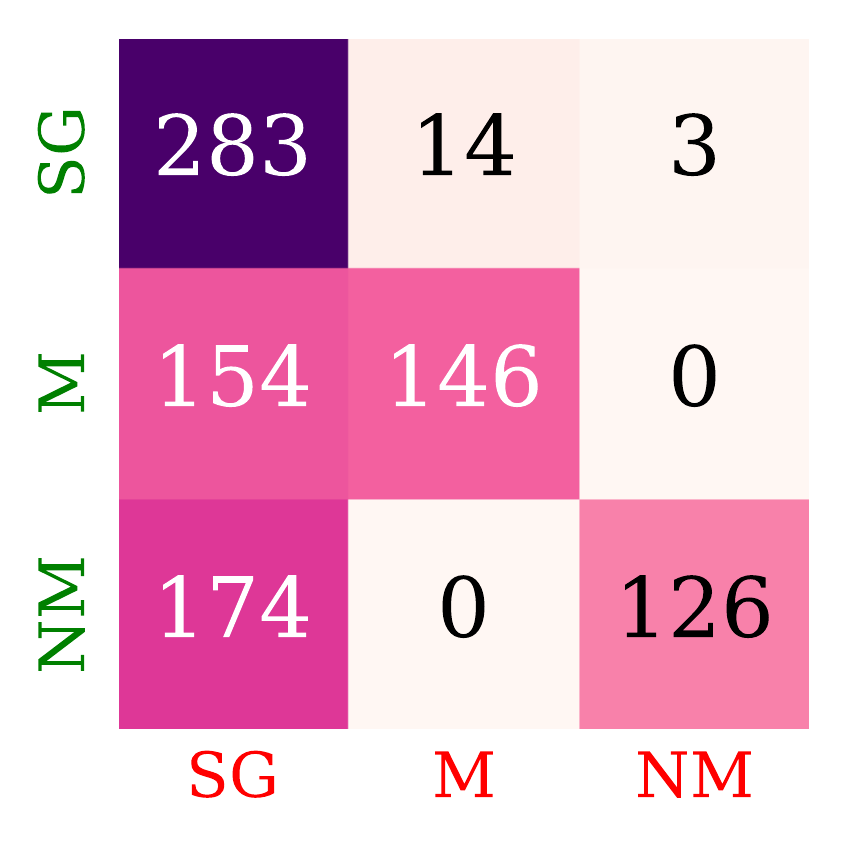} }}%
    \subfloat[\label{subfig:c2}]   
    {\includegraphics[width=4.3cm]{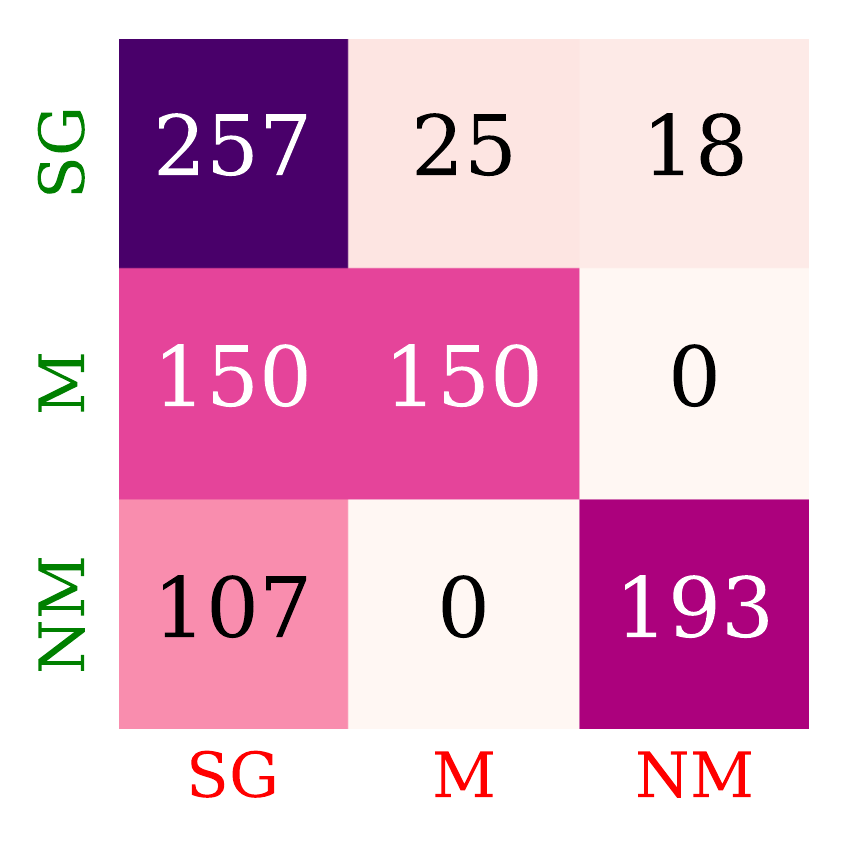} }
    \subfloat[\label{subfig:c3}]{{\includegraphics[width=4.3cm]{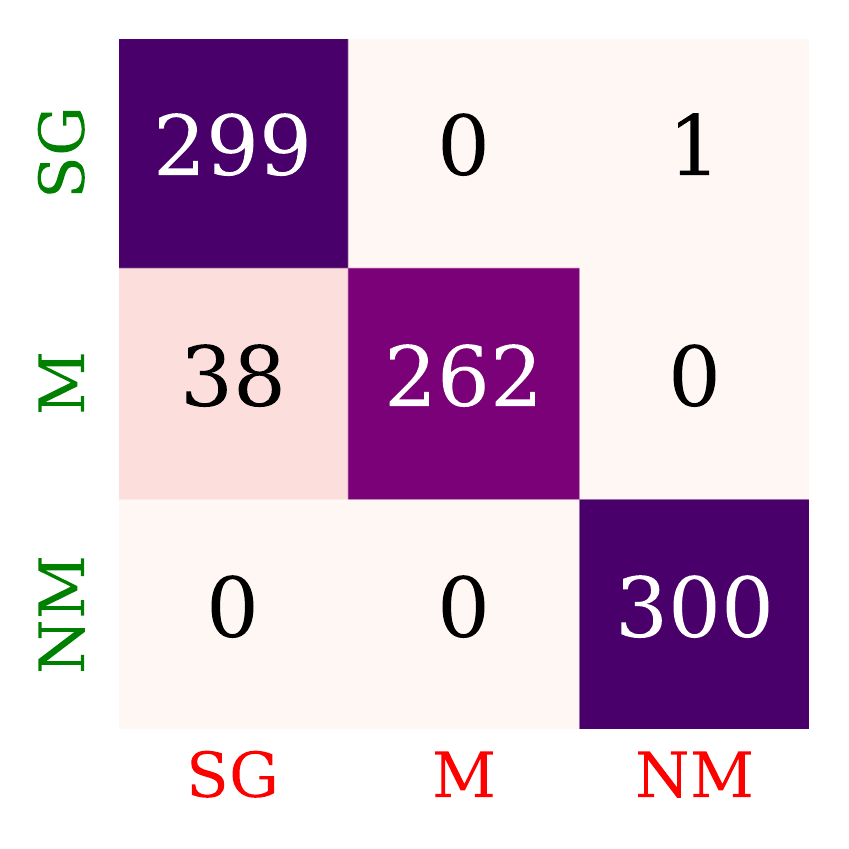} }}%
    \subfloat[ \label{subfig:c4}]{{\includegraphics[width=4.3cm]{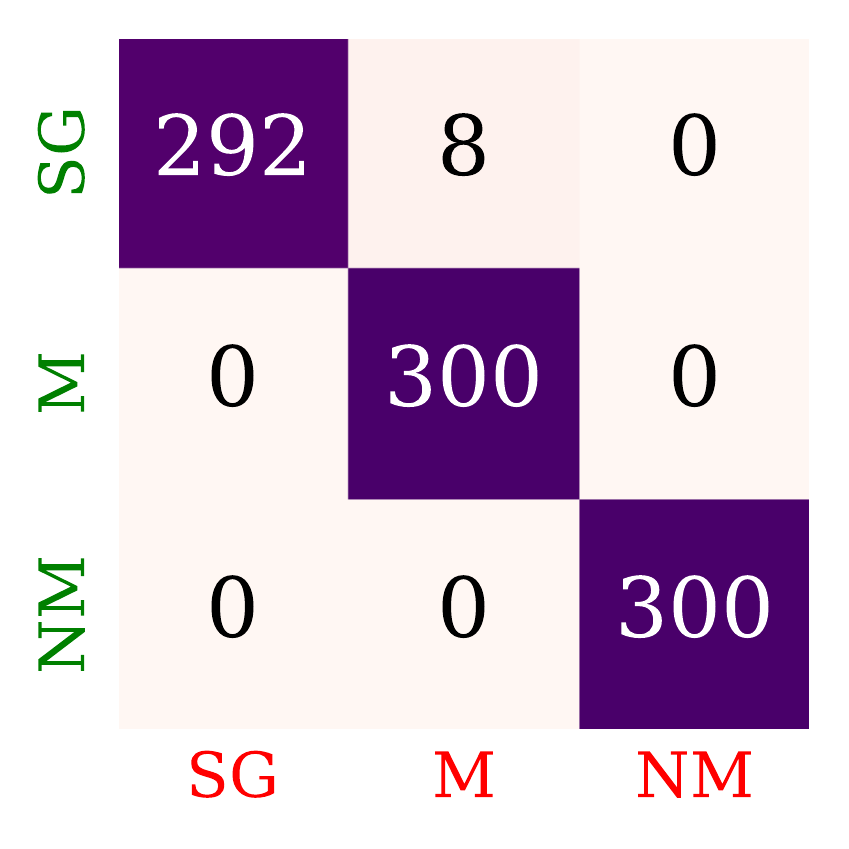} }}%
    \caption{Classification tasks involving channels not included in the training set. The considered channels are dephasing (D), dephasing in a rotated basis (D$_{\text{RB}}$), Pauli (P), Pauli in a rotated basis (P$_{\text{RB}}$), and generalized amplitude damping (GAD). Panels (a), (b), (c), and (d) depict the performance when using D as train and D$_{\text{RB}}$ as test (accuracy 62\%), D as train and P$_{\text{RB}}$ as test (accuracy 66\%), P as train and GAD as test (accuracy 96\%) and P as train and P$_{\text{RB}}$ as test (accuracy 99\%) channels, respectively. The green labels on the vertical axis represent the actual labels, while the red labels on the horizontal axis represent the predicted labels.}%
    \label{fig:allc}%
\end{figure*}

\paragraph*{Generalization Properties.---}
We evaluate the network ability to generalize, specifically its capability to correctly classify processes that were not included in the training set. We conduct four experiments to assess this: (a) Training with the standard dephasing channel D and testing on D$_{\text{RB}}$; (b) Training with D and testing on P$_{\text{RB}}$; (c) Training with P and testing on GAD; (d) Training with P and testing on P$_{\text{RB}}$. 

We present the results of these generalization tests using confusion matrices. A confusion matrix is a table used to describe the performance of a classification model, showing the actual versus predicted classifications. The confusion matrices for these experiments are depicted in \Cref{fig:allc}.

Notably, in Figure \ref{subfig:c1}, the dephasing channels do not achieve satisfactory accuracies for correctly classifying dephasing channels in a rotated basis. This is likely due to the limited information conveyed by the standard dephasing channel. However, in other scenarios, the performance improves significantly. For example, in Figure \ref{subfig:c4}, the network demonstrates the ability to correctly classify almost all Pauli channels in a rotated basis when trained with standard Pauli channels, indicating that the classification performance is largely independent of the chosen basis.

These results exhibit slight variability depending on individual runs, which can lead to better or worse outcomes in specific instances. However, in Figure \ref{subfig:c3}, the results are highly dependent on individual runs. This dependency arises when the network becomes overly specialized to the specific functions composing the decay rates, significantly impacting performance.

The results remain consistent when the full Choi matrix is considered. Additionally, to facilitate simulations, each decay rate of every process is characterized by specific functions, differing only in the values of their coefficients, randomly sampled from predefined intervals. This approach may introduce bias when attempting to generalize to other processes.

All simulations were conducted using the same network architecture and hyperparameters as those employed in the simulations shown in \Cref{fig:acc}.

\subsection{Forecasting Channel Evolution}
Forecasting in the context of time series refers to the process of predicting future values or trends based on past observations.
In this section, we focus on predicting the four diagonal components of the Choi state in the Bell basis, corresponding to either a dephasing channel in a rotated basis or a Pauli channel in a rotated basis.

The RNN is trained on historical data and its performance is assessed through the accurate prediction of future time steps. To provide a comprehensive illustration, we present the forecasting results for two representative samples: one for dephasing and one for Pauli. In each case, eight initial time steps are given, and the model is tasked with predicting the subsequent three steps.

The evaluation of our model demonstrates a remarkably low Mean Squared Error (MSE) of $10^{-5}$, indicative of the model ability to capture the dynamics of the quantum process. 

The MSE is computed as $\text{MSE} = \frac{1}{n} \sum_{i=1}^{n} \|\bm{\mathcal{C}}^{(i)} - \hat{\bm{\mathcal{C}}}{}^{(i)}\|^2 $, where \(n\) is the number of samples, \(\bm{\mathcal{C}}^{(i)}\) is a vector containing the true values of the Choi state components in the prediction time window, and $\hat{\bm{\mathcal{C}}}{}^{(i)}$ represents the vector of predicted values of the Choi state components.

\begin{figure}[]
    \centering
    \begin{subfigure}[b]{8.5cm}
        \centering
        \includegraphics[trim={0.2cm 0.6cm 0.2cm 0.0cm},clip, width=  \textwidth]{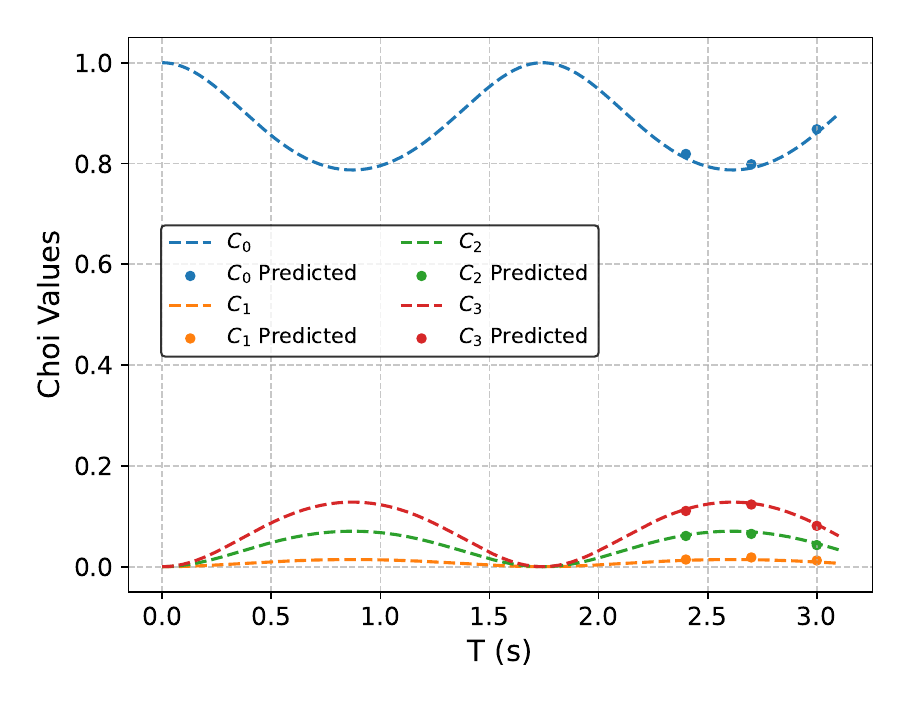}
        \caption{}
        \label{fig:subfig1}
    \end{subfigure}
    \vspace{-0.2\baselineskip} 
    \begin{subfigure}[b]{8.5cm}
        \centering
        \includegraphics[trim={0.1cm 0.6cm 0.2cm 0.0cm},clip, width=  \textwidth]{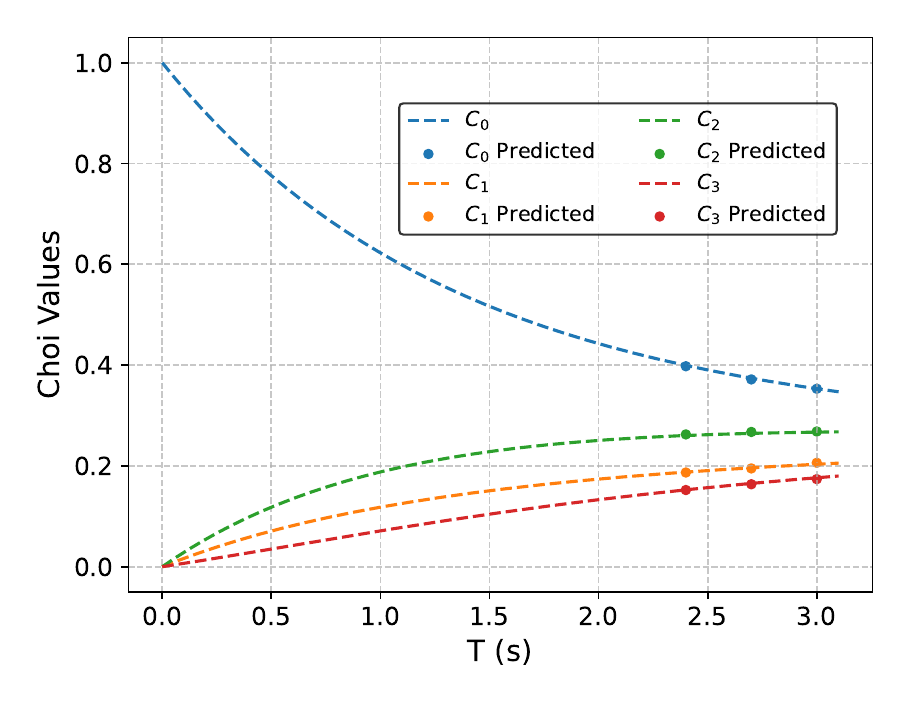}
        \caption{}
        \label{fig:subfig2}
    \end{subfigure}
    \caption{
Time series forecasting for the four diagonal elements ${C_i}$ of the Choi state in the Bell basis, for a specific test sample exposed to (a) a dephasing channel and (b) a Pauli channel in an arbitrary basis. The plot displays the actual values of $C_i$ with dashed lines, while the predicted points ($C_i$ Predicted) are shown as markers.}
    \label{fig:forecasting}
\end{figure}

The architecture comprises a GRU layer with 64 units, followed by a feedforward layer with 64 units using ReLU as activation function, and a final layer with 12 units. From \Cref{fig:forecasting} we can see that the prediction matches the actual time series.

\section{Conclusion}
\label{sec:conclusion}
This study introduces a novel methodology employing Recurrent Neural Networks (RNNs) for the classification and forecasting of Markovian and non-Markovian quantum processes based on time series data derived from Choi states. The model demonstrates exceptional accuracy, surpassing 95\% in diverse scenarios involving dephasing, Pauli, generalized amplitude damping channels, and even in the presence of noise.
The analysis extends to forecasting the evolution of Choi states, revealing the model robustness in predicting future time steps with low MSE. The utilization of only the diagonal components within the Bell basis proves effective, highlighting the efficiency of the proposed approach.\\
The findings hold implications for quantum information theory, quantum control, and quantum thermodynamics, where distinguishing between Markovian and non-Markovian dynamics is of paramount importance.\\
For future research, exploration into alternative types of data representation beyond Choi states could be pursued. For instance, investigating the feasibility of utilizing Pauli transfer matrices as time series data may provide valuable insights into the dynamics of quantum processes from a different perspective.\\
Furthermore, future research could explore higher-dimensional systems and examine the interplay between neural networks and quantum dynamics to tackle the challenges presented by more complex quantum systems and varied scenarios. For instance, upcoming work will focus on designing and utilizing real-world datasets, enabling us to rigorously evaluate the algorithms in practical settings, thereby further validating their performance and applicability in experimental contexts.\\
This exploration holds the potential to yield valuable insights into the resilience and adaptability of machine learning techniques within practical quantum computing environments.

\section{Data availability statement}
The data and code that support the findings of this study can be given on demand.

\section{Funding}
A.R.M. acknowledges support from the PNRR MUR Project PE0000023-NQSTI. 
M.F.S. acknowledges support from the PRIN MUR
Project 2022SW3RPY.
C.M. acknowledges support from the National Research Centre for HPC, Big Data and Quantum Computing, PNRR MUR Project CN0000013-ICSC. 

\appendix







\section{Master Equation from Quantum Channel}
\label{app:mefrommap}
In the pursuit of characterizing the open dynamics of quantum systems, a fundamental aspect lies in the study of the master equation governing the evolution of the system density matrix.

Our starting point is the master equation expressed as $\dot{\rho}(t) = \mathscr{L}_t[\rho(t)]$, where $\mathscr{L}_t$ is the Liouvillian superoperator. 
A corresponding set of quantum channels is generally written as $\rho(t) = \mathcal{N}_t(\rho(0)) = \sum_i E_i(t) \rho(0) E_i^\dagger(t)$. 

Assuming the invertibility of the map $\mathcal{N}_t$, one can write $\mathscr{L}_t[\cdot] = \dot{\mathcal{N}}_t\mathcal{N}_t^{-1}[\cdot]$~\cite{andersson2007finding}.
The operators $\dot{\mathcal{N}}_t$ and $\mathcal{N}_t^{-1}$ can be computed starting from an alternative representation of the channel $\mathcal{N}_t$ as $K_t = \sum_i E_i(t) \otimes \bar{E_i}(t)$~\cite{watrous2018theory}, where $\{E_i (t)\}$ are the Kraus operators. 
In this way, given a quantum state $\rho$, we can derive the Lindblad operator as $\dot{K}_tK_t^{-1}(\vec{\rho})$, where $\vec{\rho}$ is the vectorization of the density matrix $\rho$~\cite{gilchrist2009vectorization}.

\section{Recurrent Neural Networks}
\paragraph*{LSTM.---}
\label{app:rnns}
LSTM consists of a memory cell and a set of gates to regulate information flow~\cite{wang2017new}. These gates include the forget gate $f_t$, responsible for deciding which information from the cell state ${c}_t$ to discard; the input gate $i_t$, determining the information to add to the cell state; and the output gate $o_t$, controlling the information to transmit to the output $h_t$. The equations governing these operations are
\begin{align}
    f_t &= \sigma(W_f \cdot [h_{t-1},x_t] + b_f), \notag \\
    i_t &= \sigma(W_i \cdot [h_{t-1},x_t] + b_i), \notag\\
    o_t &= \sigma(W_o \cdot [h_{t-1},x_t] + b_o), \notag\\
    \Tilde{c}_t &= \tanh(W_c \cdot [h_{t-1},x_t] + b_c), \notag\\
    c_t &= f_t \cdot c_{t-1} + i_t \cdot \Tilde{c}_t, \notag\\
    h_t &= o_t \cdot \tanh(c_t).
\end{align}
Here, $\sigma(\cdot)$ represents the sigmoid function, and $W_i$ and $b_i$, with $i \in \{f, i, o, c\}$, denote the weights and biases specific to each gate. The input is represented by $x_t$ while $h_t$ is the so-called hidden state, that acts as the memory of the network. The operation $[\cdot, \cdot]$ denotes vector concatenation.
\paragraph*{GRU.---}
Being a simplified variant of LSTM, GRU merges the memory cell and the hidden state. Its fundamental components include the update gate $z_t$, responsible for governing the extent to which the past hidden state is retained, the reset gate $r_t$, determining the degree to which the past hidden state is overlooked, and the current memory content $h_t$, reflecting the updated hidden state. The main gates are~\cite{dey2017gate}:
\begin{align}
    h_t &= (1-z_t) \odot h_{t-1} + z_t \odot \Tilde{h}_t, \notag\\
    \Tilde{h}_t &= \tanh(W_hx_t+U_h(r_t \odot h_{t-1}) + b_h), \notag\\
    z_t &= \sigma(W_zx_t + U_zh_{t-1} + b_z), \notag\\
    r_t &= \sigma(W_rx_t + U_rh_{t-1}+b_r).
\end{align}
Here, $W_i, U_i$ and $b_i$ with $i \in \{ h,z,r\}$ denote the specific parameters for each gate, and $\odot$ is the Hadamard product.

\section{Data Generation}
\label{app:datageneration}
In this section, we outline the procedure for generating our dataset utilized in training the network. We emphasize that the functions chosen to model the decay rates of the master equation were selected to ensure that the integrated maps are completely positive (CP) at all times.

\paragraph*{Dephasing Channel.---} Beginning with \eqref{eq:gRB}, we observe that the parameters governing the generation of diverse instances of a dephasing channel in a rotated basis are characterized by the pair $(\theta, \varphi)$ and by $\Gamma(t) = \int_0^t \gamma(s) ds$. Here, $\cos{\theta} \in [-1, 1]$ and $\varphi \in [0, 2\pi)$ are randomly sampled from a uniform distribution for each individual sample.

To generate various semi-group instances, we sample $\gamma(t) = a$ for each process, where $a \in [0, 1)$. For the generation of Markovian processes, we choose $\gamma(t) = a\tanh t$ or $\gamma(t) =\sin^2(bt)$, with $b \in [2,4)$. Conversely, for non-Markovian samples, we employ $\gamma(t) = \sin{bt}$.

\paragraph*{Pauli Channel.---} 
The Choi state in \eqref{eq:choigp} relies on the cumulative functions $\Gamma_k(t) = \int_0^t \gamma_k(s) ds$. To generate semi-group samples, we set $\gamma_k(t) = a_k$, where $a_k \in [0, 1)$. For Markovian samples, we adopt a strategy where $\gamma_k(t) = \sin^2(b_kt)$ or $\gamma_k(t) =a_k\tanh t$, with $b_k \in [2,4)$ and $a_k \in [0,1)$. For non-Markovian samples, we use the functions $\gamma_k(t) = \sin (b_kt)$ or consider an alternative scenario characterized by $\gamma_i = \gamma_j = a$ and $\gamma_k(t) = -a\tanh t$, where $a \geq 1$ and $i,j,k$ represent any permutation of the indices 1,2 and 3.

For samples generated from the Pauli channel in a rotated basis, we assign a unique set of three Euler angles $(\alpha, \beta, \gamma)$ to each sample.

\paragraph*{GAD Channel.---} The Choi state described in \eqref{eq:gad} relies on the parameters of the GAD channel, namely $p$ and $\lambda$. To distinguish between different classes, we adopt different parameterizations for the decay rates reported in \eqref{eq:gad}, which are linked to the channel parameters via \eqref{eq:diffgad}. For semi-group samples, the solution of the differential equations with constant $\gamma_1$ and $\gamma_2$ yields $p = \frac{\gamma_1}{\gamma_1 + \gamma_2}$ and $\lambda = 1 - e^{-(\gamma_1 + \gamma_2) t}$. In the case of Markovian samples, $p$ remains constant in $[0,1]$ while $\lambda (t) = \tanh(bt)$ with $b \in [\frac{1}{2},1)$. Finally, for non-Markovian samples, $p \in [0,1]$ and $\lambda (t) = \frac{1}{2} \sin^2(bt)$ with $b \in [2,4)$.

\section{Noise Model}
\label{app:noise}
In this work, we consider a noise model inspired by the theoretical framework outlined in \cite{macchiavello2018mixed} and the experimental scenario adopted in \cite{cuevas2017experimental, ciampini2021experimental}. We focus on the case where the dominant source of noise arises from imperfections of the initial Bell state that is used at the input of the quantum channels in order to obtain the time series of the Choi states at the output.

Typically, the ideal input Bell state $\dyad{\phi^+}$ is replaced by a noisy Werner state with fidelity $F<1$, namely
\begin{equation}
\nu_F = F \dyad{\phi^+} + \frac{1-F}{3}  \sum_{i=1}^3\dyad{\phi_i} ,
\end{equation}
where $|\phi_1\rangle = |\phi^-\rangle$, $|\phi_2\rangle = |\psi^+\rangle$ and $|\phi_3\rangle = |\psi^-\rangle$ are the remaining Bell states, and $F=\langle\phi^+|\nu_F|\phi^+\rangle$.\\
Then, in this noisy scenario, the ideal Choi states of \Cref{eq:choiformula} are replaced by
\begin{equation}
    \mathcal{C}_\Lambda^F(t, t_0) = (\Lambda(t, t_0) \otimes I) \nu_F.
    \label{eq:idealbloch}
\end{equation}
Correspondingly, the diagonal elements \( r_j = \langle \phi_j |\mathcal{C}_\Lambda (t, t_0)|\phi_j\rangle \) of the Choi state in the Bell basis for ideal input are replaced by 
\begin{equation}
    s_j = \langle \phi_j | \mathcal{C}^F_\Lambda(t,t_0)|\phi_j\rangle = F r_j + \frac{1 - F}{3} \sum_{i \neq j} r_i.
    \label{eq:noisybloch}
\end{equation}
The noise model captures a realistic experimental setting where the fidelity of the input state of the channels is less than 1, due to imperfections in the preparation of the Bell state.\\
We remark that in \cite{cuevas2017experimental, ciampini2021experimental} the fidelity was experimentally estimated as $F = 0.979 \pm 0.011$.

\bibliography{bibliography}
\end{document}